\documentclass[aps,
prb,superscriptaddress,
reprint,
amsmath,
amssymb
]{revtex4-1}
\usepackage[english]{babel}
\usepackage{amsmath,graphicx,color}

\newcommand{\nocomma}{}
\newcommand{\tmop}[1]{\ensuremath{\operatorname{#1}}}

\begin{document}

\title{Quantum anomalous Hall state from spatially decaying interactions on the decorated honeycomb lattice}

\author{Mengsu Chen}
\affiliation{Department of Physics, Virginia Tech, Blacksburg, Virginia 24061, USA}
\author{Hoi-Yin Hui}
\affiliation{Department of Physics, Virginia Tech, Blacksburg, Virginia 24061, USA}
\author{Sumanta Tewari}
\affiliation{Department of Physics, Indian Institute of Technology, Kharagpur, West Bengal 721302, India}
\affiliation{Department of Physics and Astronomy, Clemson University, Clemson, South Carolina 29634, USA }
\author{V. W. Scarola}
\affiliation{Department of Physics, Virginia Tech, Blacksburg, Virginia 24061, USA}

\begin{abstract}
Topological phases typically encode topology at the level of the single particle band structure.  But a remarkable class of models shows that quantum anomalous Hall effects can be driven exclusively by interactions, while the parent noninteracting band structure is topologically trivial.  Unfortunately, these models have so far relied on interactions that do not spatially decay and are therefore unphysical.  We study a model of spinless fermions on a decorated honeycomb lattice.  Using complementary methods, mean-field theory and exact diagonalization, we find a robust quantum anomalous Hall phase arising from spatially decaying interactions.  Our finding indicates that the quantum anomalous Hall effect driven entirely by interactions is a surprisingly robust and realistic phenomenon.
\end{abstract}

{\maketitle}  
\section{Introduction}
Topological insulators (TI) are an exotic state of quantum matter distinguished from ordinary band insulators by the presence of topologically protected edge states \cite{hasan:2010,qi:2011}. While the bulk spectrum of TIs is gapped, the gapless edge (2D) and surface (3D) states remain conducting as long as time reversal symmetry (TRS)  remains unbroken. In addition to TRS-protected TIs, topological states can also occur in TRS-broken systems. The quantum anomalous Hall (QAH) insulator is one such state in which Hall resistance, defined by the voltage across the transverse direction divided by a longitudinal current, can be non-zero and quantized as a result of broken TRS \cite{nagaosa:2010}. In these cases 
the topological states of matter are characterized by a 
$\mathcal{Z}_2$ invariant (TI) and a $\mathcal{Z}$ invariant or Chern number (QAH). For non-trivial values of the invariants, the topological properties
are controlled by the band inversion phenomena and are essentially encoded at the level of the band structure, while the role of inter-particle interactions is minimal.

It has been realized that an alternative route to topological phenomena may be driven exclusively by interparticle interactions \cite{wen:1989,wen:1991}. The possibility of realizing interaction-generated topological states of matter in the charge sector -- known as topological Mott insulators \cite{raghu:2008} --even when the
non-interacting band structure is topologically trivial, can greatly expand the availability of topologically non-trivial systems, and is thus of paramount importance.

Noninteracting electronic systems in which the Fermi surface shrinks to a discrete number of Fermi points is an emerging frontier in condensed matter physics. A number of proposals have been recently put forward where QAH states are induced in mean field theory (MFT) purely by interactions in a model of band structure with Fermi points, TRS, and trivial band topology \cite{raghu:2008,wen:2010}. Such states have been proposed to appear from interactions leading to microscopic loop currents and spontaneous breakdown of TRS \cite{raghu:2008,sun:2009,zhang:2009,weeks:2010a,wen:2010,pesin:2010,liu:2010,castro:2011,sun:2011a,yang:2011,sun:2011a,fiete:2012,garcia:2013,grushin:2013,roy:2013,herbut:2014,kourtis:2014,duric:2014,tsai:2015,motruk:2015,capponi:2015,dauphin:2016,zhu:2016,wu:2016a}. However, in the most common example of Fermi points--Dirac points with emerging low energy relativistic invariance as in graphene--MFT and numerical techniques such as exact diagonalization (ED) and density matrix renormalization group fail to agree thus casting doubt on the existence of interaction-induced spontaneous QAH ground state as the true ground state \cite{raghu:2008,weeks:2010a,nishimoto:2010,garcia:2013,grushin:2013,roy:2013,roy:2014,duric:2014, pollmann:2014,motruk:2015,capponi:2015,kurita:2016}.   But according to these calculations, the key condition for the emergence of the QAH state (if at all) is that second and/or third nearest neighbor repulsive interactions need to be equal to or larger than nearest-neighbor repulsion, or having long-range hopping with direction-dependent phases in the hopping amplitudes \cite{sun:2009,wu:2016a}. Both of these conditions are difficult to achieve  [Ref.~\onlinecite{liu:2016} proposes that the Ruderman-Kittel-Kasuya-Yosida (RKKY) interaction could offer a workaround for the former condition].  

\begin{figure}[h] 
  \includegraphics[width=1.0\columnwidth]
  {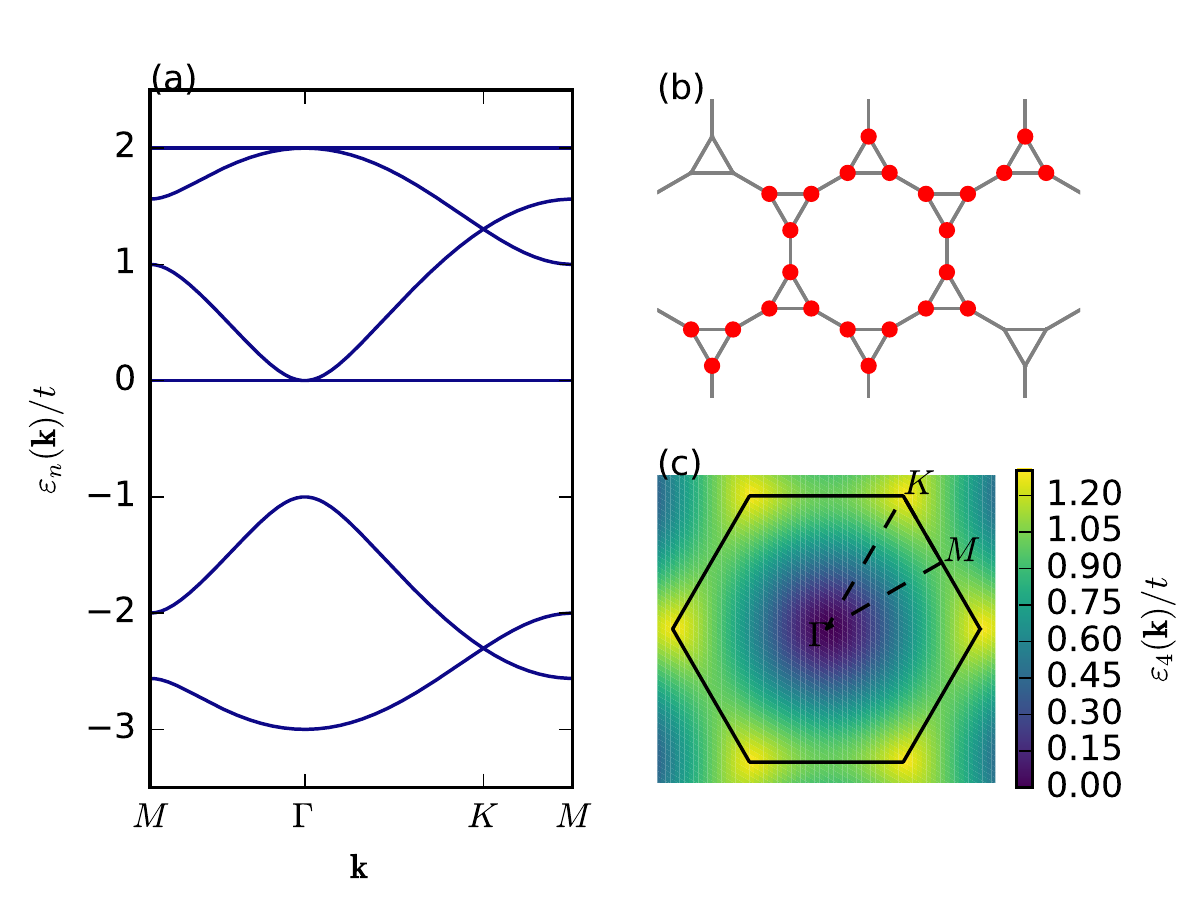}
  \caption{(a) Band structure along the high-symmetry path $M\to\Gamma\to K \to M$ [dashed line in Fig. (c)] in the first Brillouin zone. (b) The decorated honeycomb lattice and the 24-site cluster (red dots) studied by exact diagonalization. (c) The first Brillouin zone of the decorated honeycomb lattice and the band (color map) right above the Fermi point (a QBCP).
  \label{fig_band}}
\end{figure}

Most MFT studies find that non-spatially decaying interactions are important for establishing the QAH phase but there are exceptions.  Studies of 
spinless fermions on certain lattices with a quadratic band crossing point (QBCP) as the non-interacting Fermi surface offer a different starting point \cite{sun:2009,wen:2010,zhang:2009,tsai:2015}.  One example, mentioned above, is the checkerboard lattice \cite{sun:2009}.  An ED study \cite{wu:2016a} found a QAH phase arising from spatially decaying interactions but anisotropic hopping and equal hopping amplitude for different sublattices were needed to stabilize the phase \cite{wu:2016a}. MFT studies on a different lattice with a QBCP, a decorated honeycomb (or star) lattice, indicate that spatially decaying interactions might support a robust QAH phase \cite{wen:2010} even with just nearest-neighbor interactions.  But MFT is notoriously susceptible to quantum fluctuations that can significantly impact phase diagrams. Therefore the realizability of the QAH phase from spatially decaying interactions on the decorated honeycomb lattice remains an open issue.

In this paper, we investigate the prospect of generating
interaction-driven QAH states on a decorated honeycomb lattice at half filling.  Our principal objective is to investigate if the QAH state occurs for physically realizable interaction parameters that can be used to model realistic  systems, e.g., electrons in solids or dipolar fermions in optical lattices. We present ED results which incorporate quantum fluctuations but are applicable only to small system sizes, as well as mean-field results \cite{wen:2010} which are approximate (could not capture quantum fluctuations) but can handle the system in the thermodynamic limit.   The complementary nature of the two approaches, coupled with qualitatively very similar phase diagrams obtained from both of them, gives us confidence about the reliability of our calculations. Remarkably, in both ED and MFT, we find that a QAH state occurs on the decorated honeycomb lattice at the QBCP, for interaction parameters that \textit{progressively decrease with separation}.  The agreement between MFT and ED implies that quantum fluctuations allow a robust QAH phase in this lattice.  Our work sets the stage for observations of QAH under realistic conditions of spatially decaying interactions.

The paper is organized as follows. In Sec.~\ref{sec_model} we present the model of interacting fermions on a decorated honeycomb lattice.  We also review the complementary methods used to solve the model: numerical ED and a MFT obtained from the Hartree-Fock approximation.  In Section~\ref{sec_phase_diagram} we present the phase diagram obtained from both methods.  Here we show that both methods agree at weak to moderate interaction strengths.  Sec.~\ref{sec_qah} and \ref{sec_spatial} discuss, respectively, the uniform QAH phase and the CDWs found from both methods. We summarize by discussing possible connections to materials and ultracold atom gas experiments in Sec.~\ref{sec_summary}.

\begin{figure}
  \resizebox{.8\columnwidth}{!}{\includegraphics{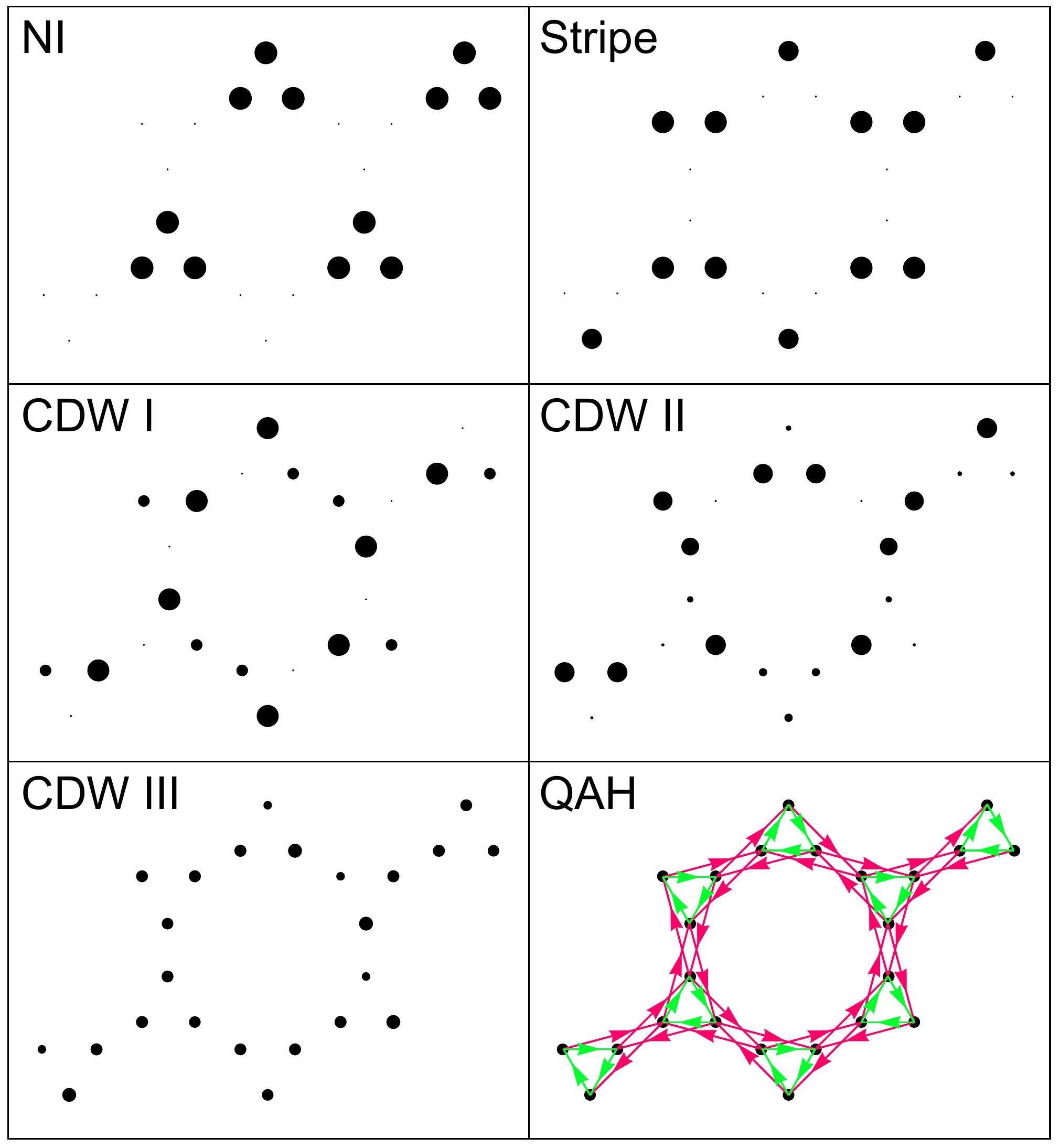}}
  \caption{Density patterns for the Nematic Insulator (NI), Stripe, and Charge Density Waves (CDWs).  The QAH pattern draws bond currents computed with arrows indicating the direction of the currents $\tmop{Im}
  \langle c_i^{\dagger} c_j \rangle$.  All plots are the result of mean field calculations but exact diagonalization produced the same configurations where comparisons could be made (see Appendices~\ref{App:ED} and~\ref{App:corr}).    \label{mf_pattern}}
\end{figure}

 \begin{figure}[t]
  \resizebox{.9\columnwidth}{!}{\includegraphics{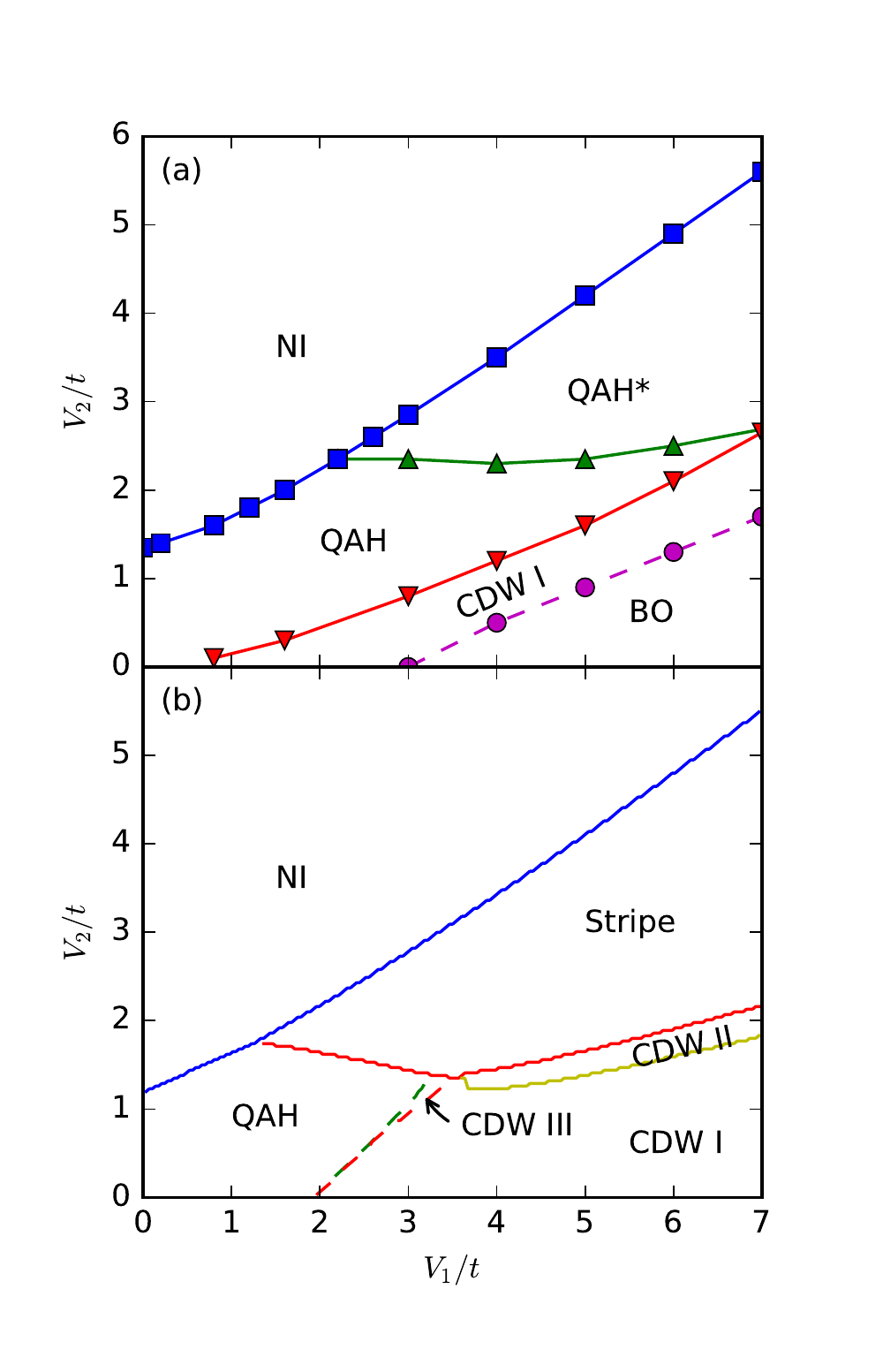}}
  \caption{(a) Phase diagram obtained using exact diagonalization on Eq.~\eqref{eq_H}.  The symbols are results from calculations and the lines are a guide to the eye.  The bond ordered (BO) phase is a uniform phase that results from the superposition of bond-ordered crystal configurations (see Appendix~\ref{App:BO}).
  (b) The same as (a) but the lines plot transitions obtained from self-consistent mean field theory on an infinite lattice. The dashed lines indicate second order phase transitions.  The agreement between panels (a) and (b) shows strong evidence for a robust QAH phase.  The QAH order appears to survive in non-interacting limits near the origin but here the QAH gap vanishes asymptotically with interaction strength. 
  }
  \label{pd_mf}
\end{figure}

\section{Model and Methods}  
\label{sec_model}
We consider a tight-binding Hamiltonian,
\begin{equation}
  H = - t \sum_{\langle i, j \rangle} c_i^{\dagger} c_j + V_1 \sum_{\langle i,
  j \rangle} n_i n_j + V_2 \sum_{\langle \langle i, j \rangle \rangle} n_i n_j,\label{eq_H}
\end{equation}
where $c_i^{\dagger}$ creates a spinless fermion on site $i$.
The summation is over nearest- ($\langle i,j \rangle$) or
next-nearest- ($\langle\langle i,j \rangle\rangle$) neighbors
on a decorated honeycomb lattice [Fig.~\ref{fig_band}(b)].  Henceforth we consider $V_1,V_2\geq0$ and set $t=1$ as the energy scale.

Fig.~\ref{fig_band}(a) shows the non-interacting band structure of Eq.~\eqref{eq_H}.  At $1/2$ or $5/6$ filling ratios, the 
Fermi surface shows a QBCP at the $\Gamma$ point, which 
is an important component to realizing a QAH ground state.  For definiteness we focus on half-filling.  

\begin{figure}
  \resizebox{.8\columnwidth}{!}{\includegraphics{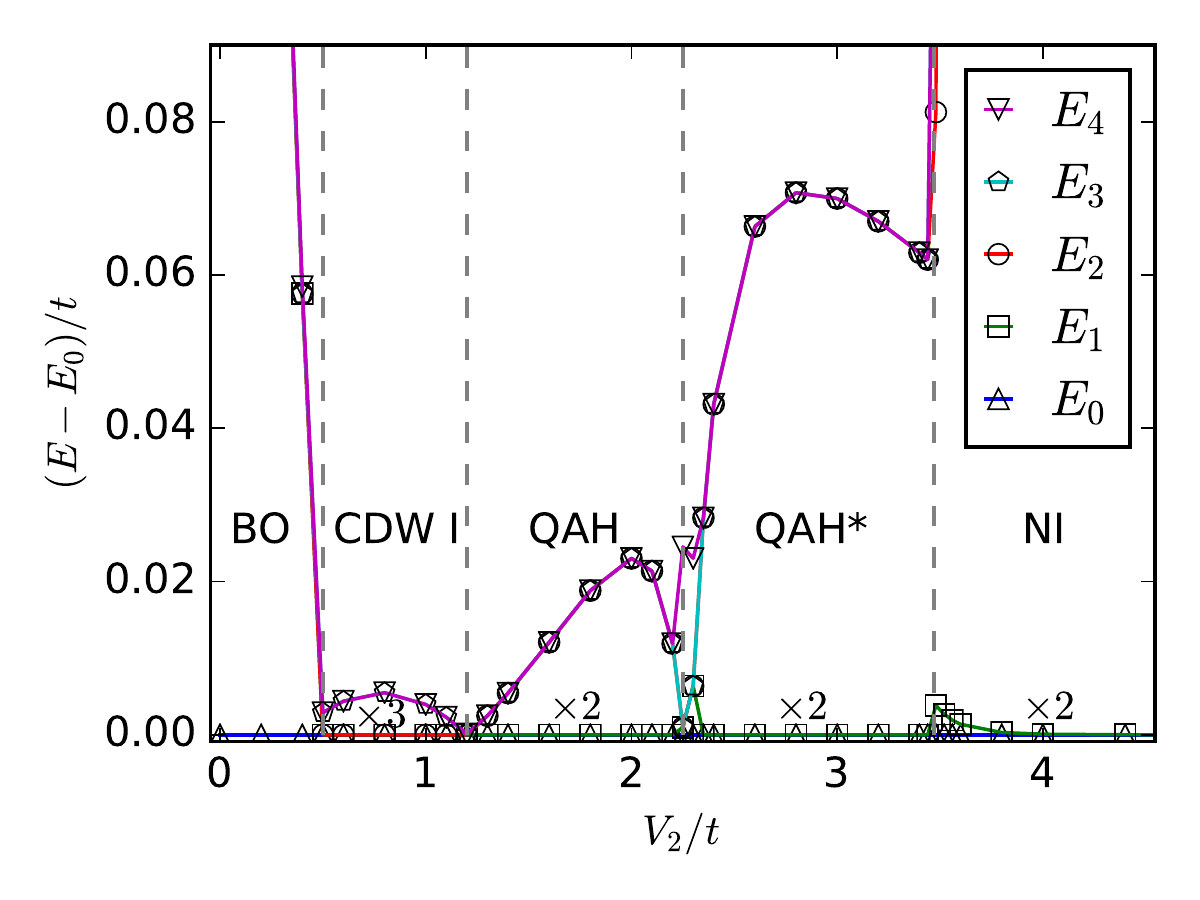}}
  \caption{Exact diagonalization calculation of the lowest 5 energies as a function of $V_2$ for $V_1=4$.  The vertical dashed lines mark phase transitions determined by level crossings and order parameters. The numbers after '$\times$' label the ground state degeneracy.
 \label{figEgap}}
\end{figure}

To study the phase diagram where the hopping strength is on the order of the interaction strength we use two complementary methods.  We employ a modified Arnolidi algorithm (the Krylov-Schur algorithm \cite{stewart:2002}) which allows study of large sparse matrices with degenerate eigenvalues.  We refer to application of this method as ED because it is unbiased and gives the same results as other unbiased methods on small lattices.  With ED we work on a finite system size ($N_e = 12$ fermions on $24$ sites, i.e., $2 \times 2$ unit cells).  ED yields the lowest energy states and includes all interaction terms without approximation. 

We use periodic boundaries. Periodic boundary conditions are widely used to study bulk properties by eliminating boundary effects \cite{capponi:2015, wu:2016a, zhu:2016}. Prior work has found that boundary effects could produce level crossings that are not related to topological transitions \cite{capponi:2015}.  We use periodic boundary conditions to avoid possible artificial phase transitions and minimize finite size effects. 
 
We also employ MFT on an infinite lattice to complement the results of ED which is limited to a small lattice.  MFT first decouples the interaction terms as:
\begin{eqnarray}
n_{i}n_{j} & \rightarrow & n_{i}\left\langle n_{j}\right\rangle +\left\langle n_{i}\right\rangle n_{j}-\left\langle n_i\right\rangle \left\langle n_j\right\rangle -c_i^{\dagger}c_j\left\langle c_j^{\dagger}c_i\right\rangle \nonumber \\
 &  & -\left\langle c_i^{\dagger}c_j\right\rangle c_j^{\dagger}c_i+\left\langle c_i^{\dagger}c_j\right\rangle \left\langle c_j^{\dagger}c_i\right\rangle ,\label{eq:HFdecouple}
\end{eqnarray}
and then follows with a self-consistent computation of $\langle c_i^\dagger c_j\rangle$ with the same unit cell as ED but on an infinite lattice (see Appendix~\ref{App:MFT} for details).
With our chosen unit cell of 24 sites, there are 108 independent values of $\langle c_i^\dagger c_j\rangle$ which need to be self-consistently solved.

\section{Phase diagram}
\label{sec_phase_diagram}

Extreme limits of the phase diagram can be computed very precisely.
For infinitely large interaction strengths the phases of  Eq.~\eqref{eq_H} are classical states that spontaneously break spatial lattice symmetries.  For $V_1\rightarrow\infty$ a degenerate manifold of classical states that avoid nearest neighbor pairs forms.  For $V_2\rightarrow\infty$ the lowest energy ground state at half filling is a nematic insulator (NI) which avoids forming next-nearest neighbor pairs.  
 For $V_1\sim V_2\rightarrow\infty$ a stripe phase sets in.  Fig.~\ref{mf_pattern} depicts the charge configurations of the NI and stripe phases.  We have checked that both ED and MFT give the same ground states for infinitely large interaction strengths.  
These and other classical phases compete with the uniform QAH phase as the interaction energies are lowered to be on the order of $t$.

We expect quantum phases when energy scales compete, i.e., $t\sim V_1\sim V_2$. 
Fig.~\ref{pd_mf} shows our central result, the phase diagrams obtained by both ED and MFT in this regime.   We see qualitative agreement between both methods in several parts of the phase diagram.  Most importantly, we see that the QAH phase arises for nearly the same parameters in both ED and MFT. For example, in the limit of $V_1\rightarrow 0$, both ED and MFT find QAH for $V_2<1.2t$ and NI for $V_2>1.2t$. We now discuss important aspects of these phases and our method for obtaining the phase boundaries.

 \begin{figure}[t]
  \includegraphics[width=.9\columnwidth]{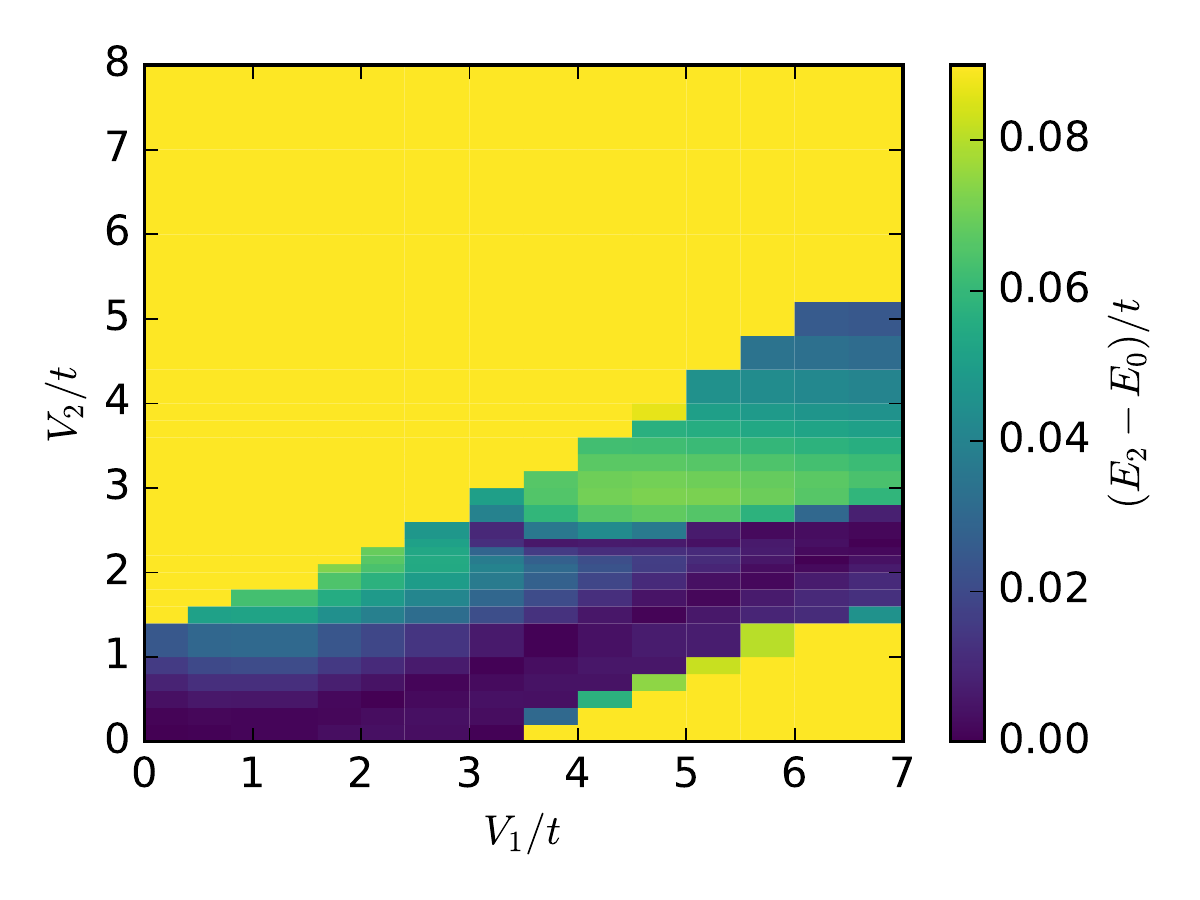}
  \caption{Energy gap, $E_2-E_0$, obtained from exact diagonalization, where $E_0$ is the lowest energy level, and $E_2$ is the third lowest level.  (The second level, $E_1$, is always degenerate with $E_0$ within the QAH phase.)  The gapped QAH and QAH* phases are separated by a nearly horizontal gap-closing line. The QAH gap also closes along the axes. \label{fig_E2-E0}}
\end{figure}

\section{Quantum anomalous Hall phase}
\label{sec_qah}

The QAH phase has uniform density and is gapped in the bulk.  It can carry current along its edge.  Although the ground state is two-fold degenerate (one state for each direction of current flow) it spontaneously breaks TRS so that only one direction of edge current is chosen for arbitrarily small perturbations.  The QAH state is therefore a topological phase with a definite Chern number and should reveal quantized Hall resistance.  

At small (but non-zero) interaction strengths, both ED and MFT predict the existence of a QAH state.  Since these methods make distinct approximations, we take the qualitative agreement between the results as definitive evidence for the presence of a QAH state (even for decaying interactions, $V_1>V_2$).  To find the QAH phase we compute local currents:
\begin{eqnarray}
J_{ij}=i (\langle c_i^{\dagger} c_j\rangle -\langle c_j^{\dagger} c_i\rangle),
\end{eqnarray}
that can be non-zero only in a phase with spontaneously broken TRS.  Fig.~\ref{mf_pattern} shows the current pattern in the QAH phase found using both methods.

At first glance, ED should not be able to detect QAH phase since each of the two ground states obtained numerically are
arbitrary superpositions of the time reversal pair of QAH states.   To reveal broken TRS we use the two degenerate ground states to  
compute a loop current in ED: 
$\hat{J} = \sum_{i, j \in \tmop{loop}} \hat{J}_{i \nocomma j}$,
where the summation is over a closed loop of sites, e.g., a small triangle. We then 
compute the local current pattern that maximizes $\langle \hat{J}\rangle$. (see Appendix~\ref{App:ED})

Interestingly, ED and MFT disagree at moderate values of the interaction.
Fig.~\ref{figEgap} shows that ED finds a second QAH phase (QAH*) separated from QAH by a gap closing indicative of a phase transition.  
This phase is absent in the MFT calculations.  The QAH* phase is characterized by the same current pattern but in the opposite direction.  Fig.~\ref{fig_E2-E0} shows ED results for the energy difference between one of the degenerate ground states and the next highest energy state.  Here we see that the QAH gap vanishes as the interactions (either $V_1$ or $V_2$) vanish near the origin and at the transition between QAH and QAH*.

The existence of a gap and chiral currents in the QAH phases implies that the Chern number is well defined in both MFT and ED.  In the QAH phase we can rely on a finite angular momentum in the gapped ground state to define a Chern number directly because the QAH state is adiabatically connected to the non-interacting limit \cite{fang:2012,wen:2010}.

\section{Spontaneous spatial-symmetry breaking phases}
\label{sec_spatial}

Moving parameters away from the QAH regime, we find topological phase transitions to conventional phases that spontaneously break spatial symmetries.  We detect these phases using  density and density-density correlation functions to reveal long-range spatial order in the density.  We find that the current vanishes in all density-ordered phases.

 To capture the transition from the uniform QAH liquid to the spontaneous spatial-symmetry breaking phases we define the maximum density difference, $\delta n=\max_i\langle n_i\rangle-\min_i\langle n_i\rangle$, which detects density modulation.  The phase boundaries between CDW phases, and that of the QAH uniform phase,
can be identified as a sudden jump in $\delta n$ (see Appendix~\ref{App:ED}).  The dashed lines in the bottom panel of Fig.~\ref{pd_mf} show second order transitions found within MFT.   In the case of ED with degenerate ground states,
$\delta n$ is computed as the difference between the maximum and minimum eigenvalues of
the matrix of a single-site number operator on the ground state subspace. Similar transitions were found with ED. In particular, the QAH-NI phase boundary computed using ED and MFT gives very close agreement.

We also use density-density correlations, $\left<n_in_j\right>-\left<n_i\right>\left<n_j\right>$, to identify transitions, particularly to the NI phase (see Appendix~\ref{App:BO}).  In the NI phase the energy spectrum obtained from ED shows a two-fold degenerate ground state. The two-fold degeneracy originates from the symmetry breaking of $C_6$ rotation symmetry in the Hamiltonian down to the $C_3$ symmetry in the NI ground state.  

At large interactions strengths ($V_1\sim V_2\gg1$) both ED and MFT show a stripe phase. The stripe phase breaks translation symmetry by doubling the unit cell along the direction of one primitive vector. It also breaks $C_6$ rotational symmetry to $C_2$. Therefore, the ground state degeneracy should be $2\times 3=6$ fold. The energy spectrum from ED does show a 6-fold degeneracy and a stripe pattern in the density but the comparison between MFT and ED breaks down for $V_1\sim V_2\sim 1$, as shown in Fig.~\ref{pd_mf}.

There are differences between the MFT and ED phase diagrams.  ED imposes finite size effects, while MFT ignores precise treatment of quantum fluctuations.  As a result of these differences MFT favors a stripe phase while ED finds the QAH* phase for $V_1\sim V_2\sim3$.  Both methods return to agreement for   $V_1\sim V_2 \gg 1$.   

The most striking difference arises for $V_2\rightarrow0$ and $V_1\sim 5$.  Here MFT predicts a CDW I phase while ED finds a BO phase.  In the BO regime we find a uniform state with excitations that superpose a particle and hole along certain bonds of the hexagon separately.  The resulting excitations are BO crystals.  But in the ground state the hopping superposes these crystals to form a uniform state, the BO phase (see Appendix~\ref{App:BO}).  Further work will include quantum fluctuations above MFT to see if the BO phase is favored over the CDW I in the thermodynamic limit.   For  $V_2\rightarrow0$ and $V_1\rightarrow\infty$ the BO gap vanished and the ED ground states becomes massively degenerate at an energy that agrees with MFT. 

\section{Summary}
\label{sec_summary}

Topological phases typically encode effects of topology at the level of band structure through the band inversion phenomena. The effects of interactions in topological insulators are usually considered to be minimal. Here we instead consider spinless fermions on the decorated honeycomb lattice with the Fermi surface tuned to a quadratic band touching point which is topologically trivial. Using the complementary methods of MFT and ED we show that the QBCP on the decorated honeycomb lattice is unstable to short range interactions and produces a QAH phase with spontaneously broken TRS driven exclusively by interactions. The interaction-driven QAH phase survives quantum fluctuations while arising from spatially decaying interactions. 

The model considered here can serve as a toy model for certain systems. We note that this lattice structure has been realized in Iron(III) acetate \cite{Zheng:2007} and discussed in the context of ${\mathrm{Mo}}_{3}{\mathrm{S}}_{7}{(\mathrm{dmit})}_{3}$\cite{rosalusar:2004,jacko:2017}. Electrons in these materials can, in principle, be polarized to be spinless fermions discussed in this work.  Evaluation of a possible mechanism of polarization (e.g., $g$-factor engineering or ferromagnetic exchange coupling) is beyond the scope of the present work. The decorated honeycomb lattice can also be generated in optical lattices via three groups of lasers intersecting at equal angles, with each group containing three frequencies.

Spatially decaying interactions are an important criterion for realistic modeling with electrons in solids or dipoles in optical lattices.  Our results therefore set the stage for realizing interaction-driven QAH and related topological states such as chiral spin liquids \cite{wen:1991,fiete:2012} in optical lattices with physically realizable spatially decaying interactions.

\begin{acknowledgments}  VWS acknowledges support from the AFOSR (FA9550-11-1-0313) and ARO (W911NF-16-1-0182). ST acknowledges support from ARO (W911NF-16-1-0182). \end{acknowledgments}

\appendix

\section{Mean-field Theory} \label{App:MFT}

The goal of the MFT is to approximate the many-body state by a state with non-interacting particles. Consider the set of all possible states formed by non-interacting fermions only. To compute the energy $\langle H\rangle$, we need to compute the expectation value of the interaction terms, each with four operators. Since non-interacting fermions are Gaussian states, Wick's theorem can be used to compute those four-operator terms, which leads to the decomposition presented in Eq.~\ref{eq:HFdecouple} in the main text.
  
The self-consistent approach we adopt initializes the correlations $\langle c_i^\dagger c_j\rangle$ randomly. Then, at each iteration, the correlations are updated with the ground state of the resulting Hamiltonian. This process is iterated until convergence.  
  
   \begin{figure}[t]
  \resizebox{.8\columnwidth}{!}{\includegraphics{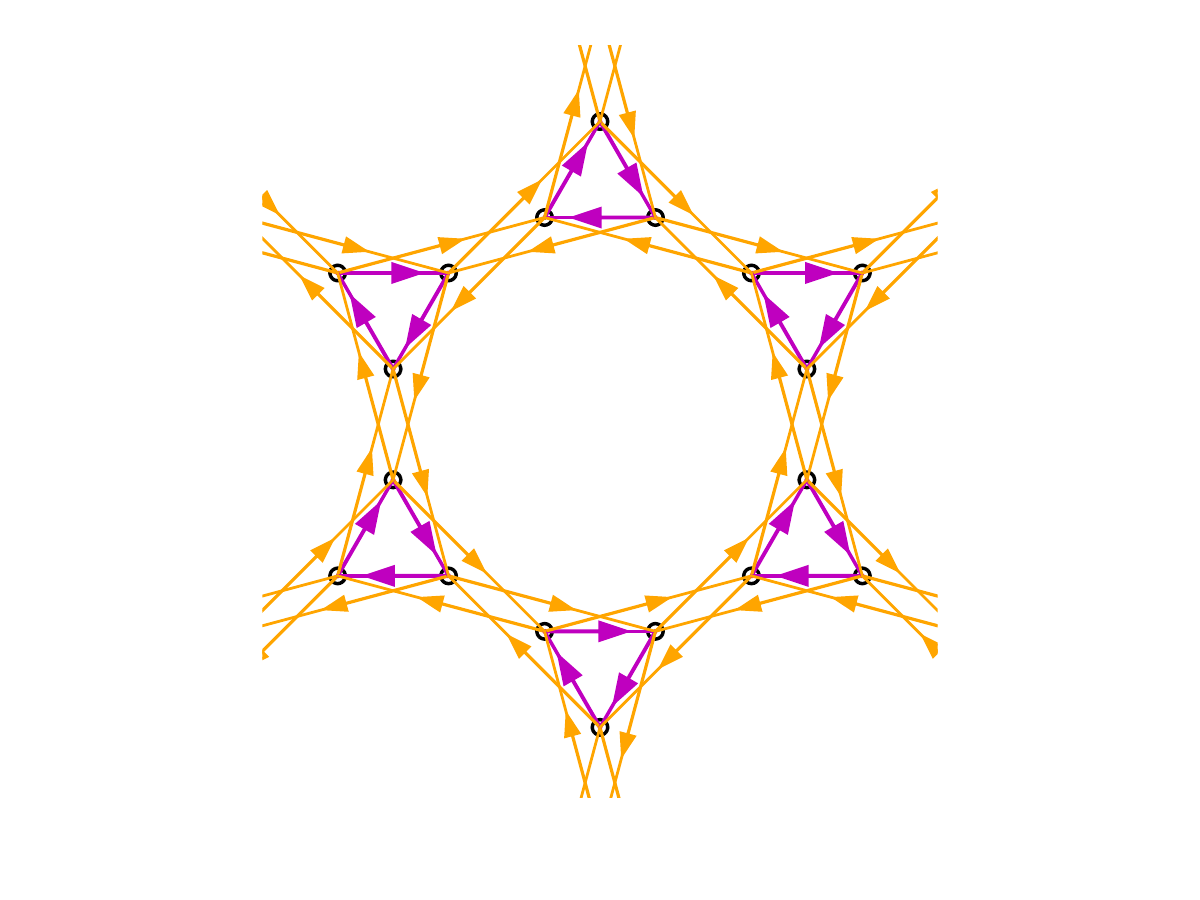}}
  \caption{Bond currents computed from exact diagonalization on one of the two QAH states at $V_1=4$ and $V_2=3$. Arrows indicate the direction of the bond currents $\tmop{Im} \langle c_i^{\dagger} c_j \rangle$, in agreement with mean field results plotted in Fig.~2 of the main text.  \label{patterns}}
\end{figure}

\section{Currents and density calculations within ED}\label{App:ED}

Within ED ground states are often degenerate.  Within the QAH phase 
arbitrary superpositions of the time reversal pairs of QAH states appear to prevent us from directly computing currents.  But we can compute current by solving for a superposition of the two degenerate ground states which maximizes the expectation value of the current operator.   

We consider the current $J_{\alpha\beta}=\langle\Psi_\alpha|\hat{J}|\Psi_\beta\rangle$ and 
denote by $| \Psi_{A/B} \rangle$ the pair of degenerate ground states.  In this subspace we compute the $2\times2$ matrix $J_{\alpha\beta}$.
Since $\hat{J}$ anticommutes with the time-reversal operator ${\cal T}$ and
${\cal T}|\Psi_{A}\rangle=|\Psi_{B}\rangle$, it is straightforward to show that
the two eigenvalues of $J_{\alpha\beta}$ are $\pm\lambda_J$ and their eigenstates are also
time-reversal pairs. Taking one of the eigenstates and computing is current pattern 
and magnitude, we find qualitative agreement with that obtained from MFT.  Fig.~\ref{patterns} shows the results from an ED calculation of the current patterns  for the QAH phase which agrees with MFT patterns presented in Fig.~2 of the main text.

 \begin{figure}[t]
  \resizebox{.8\columnwidth}{!}{\includegraphics{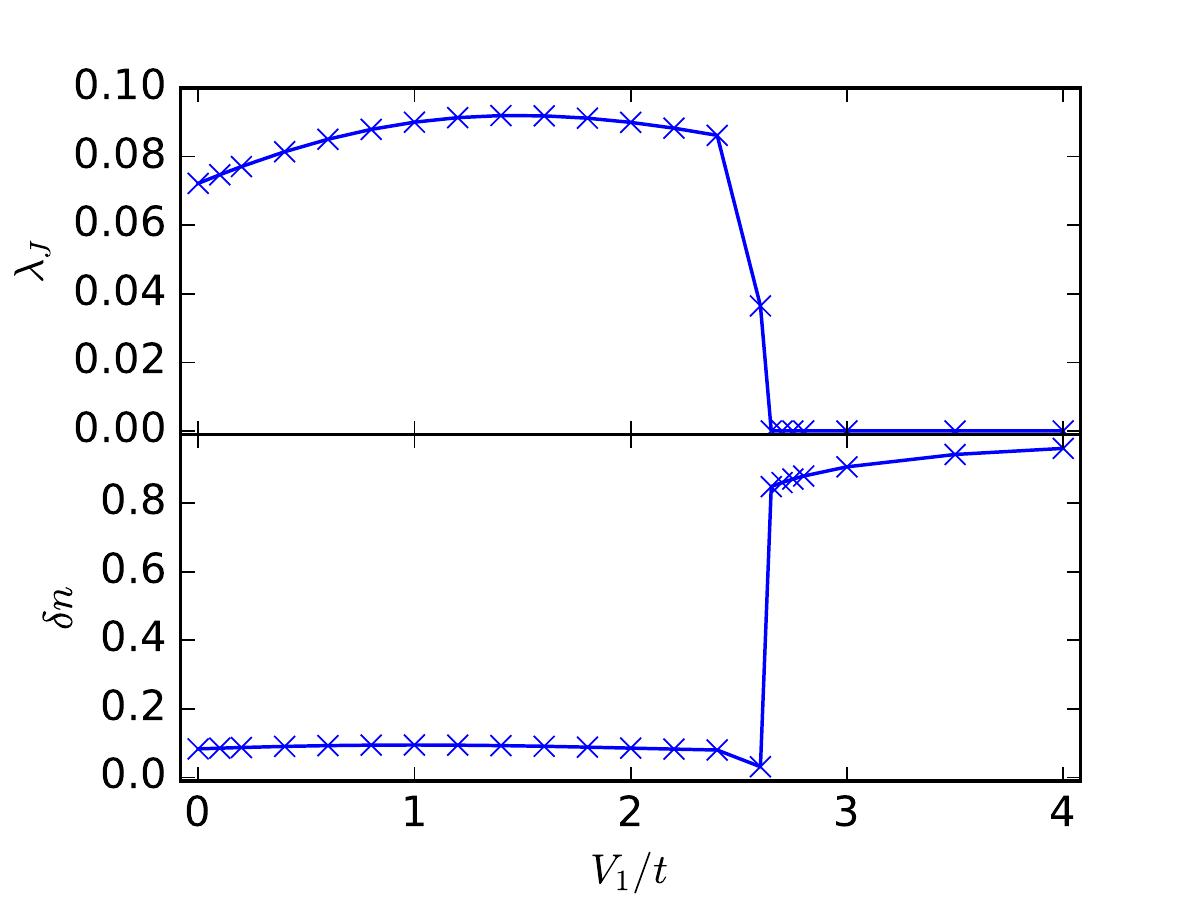}}
  \caption{Top: Exact diagonalization calculation of the QAH order parameter measuring loop current as a function of interactions strength with $V_1 = V_2$. The sudden vanishing of $\lambda_J$ indicates a transition from QAH ($\lambda_J\neq0$) to CDW ($\lambda_J=0$) along the $V_1 = V_2$ line in the phase diagram.  Bottom: The same but for the CDW order parameter measuring the maximum density difference to show a phase transition from QAH (uniform, $\delta n=0$) to CDW (nonuniform, $\delta n>0$) along
  the $V_1 = V_2$ line in the phase diagram.
  \label{figlambda}}
\end{figure}

 To demonstrate the adequacy of $\lambda_J$ as an order parameter in the QAH phase,
the top panel of Fig.~\ref{figlambda} plots $\lambda_J$ against $V_1$ along the $V_1=V_2$ line.  We 
see a sharp transition from finite and stable values of $\lambda_J$ to $\lambda_J=0$
at around $V_1=V_2\approx2.7$, indicative of a phase transition.

In the case of ED with multiple degenerate ground states we can apply a similar method to compute
$\delta n$.  It is computed as the difference between the maximum and minimum eigenvalues of
the matrix of a single-site number operator acting on the ground state subspace, similar to the procedure of computing $\lambda_J$. As an example, we plot this order parameter along a transition line from QAH to NI in the bottom panel of Fig.~\ref{figlambda}. The jump in the order parameter is at the same location as the jump in QAH order parameter, consistent with the QAH-NI transition. On the QAH side of the transition, we also find nonzero bond current for the eigenvectors of the single-site number operator on ground state subspace, and uniform density for the eigenvectors of current operator on ground state subspace. This indicates the stability of QAH under the time reversal invariant perturbation.

\begin{figure}[t]
    \includegraphics[width=.5\columnwidth]{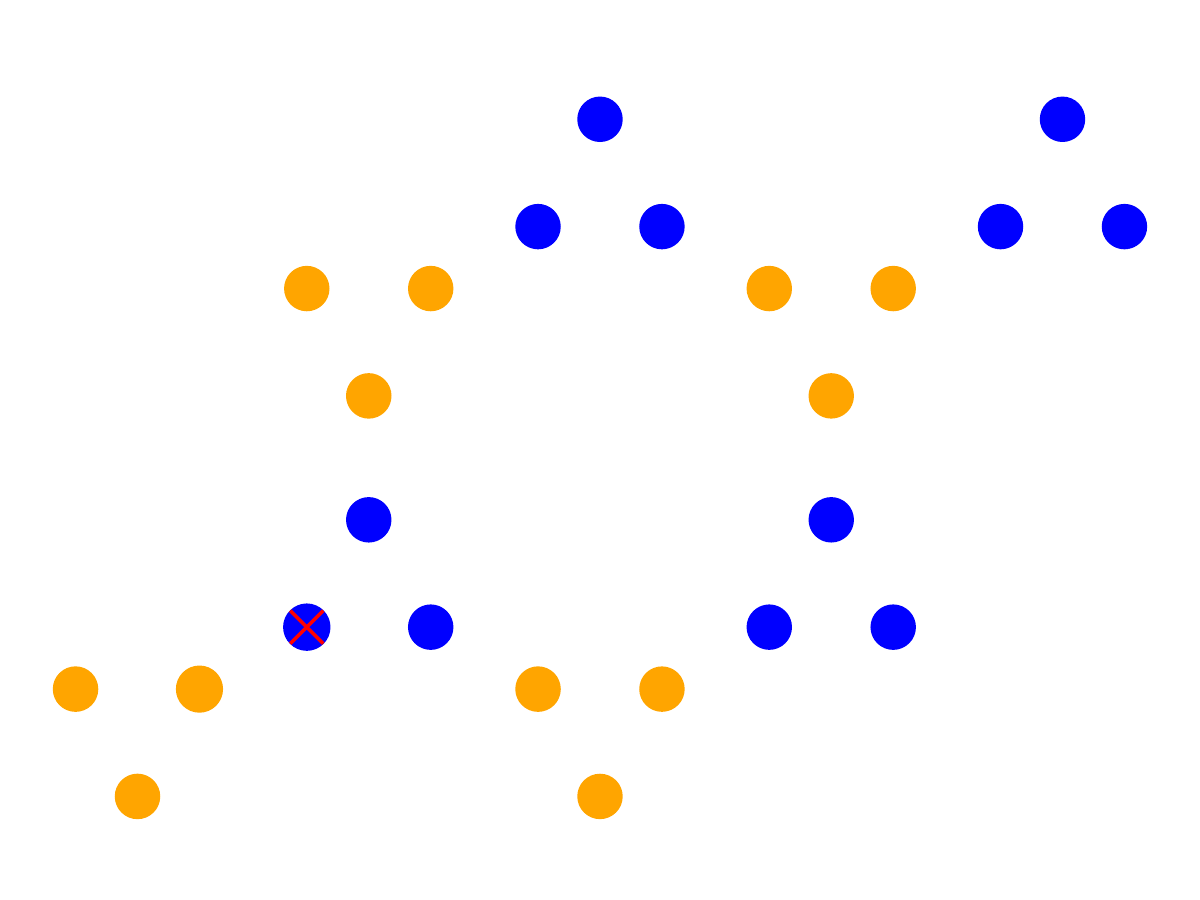}
    \hspace{-.1in}
    \includegraphics[width=.5\columnwidth]{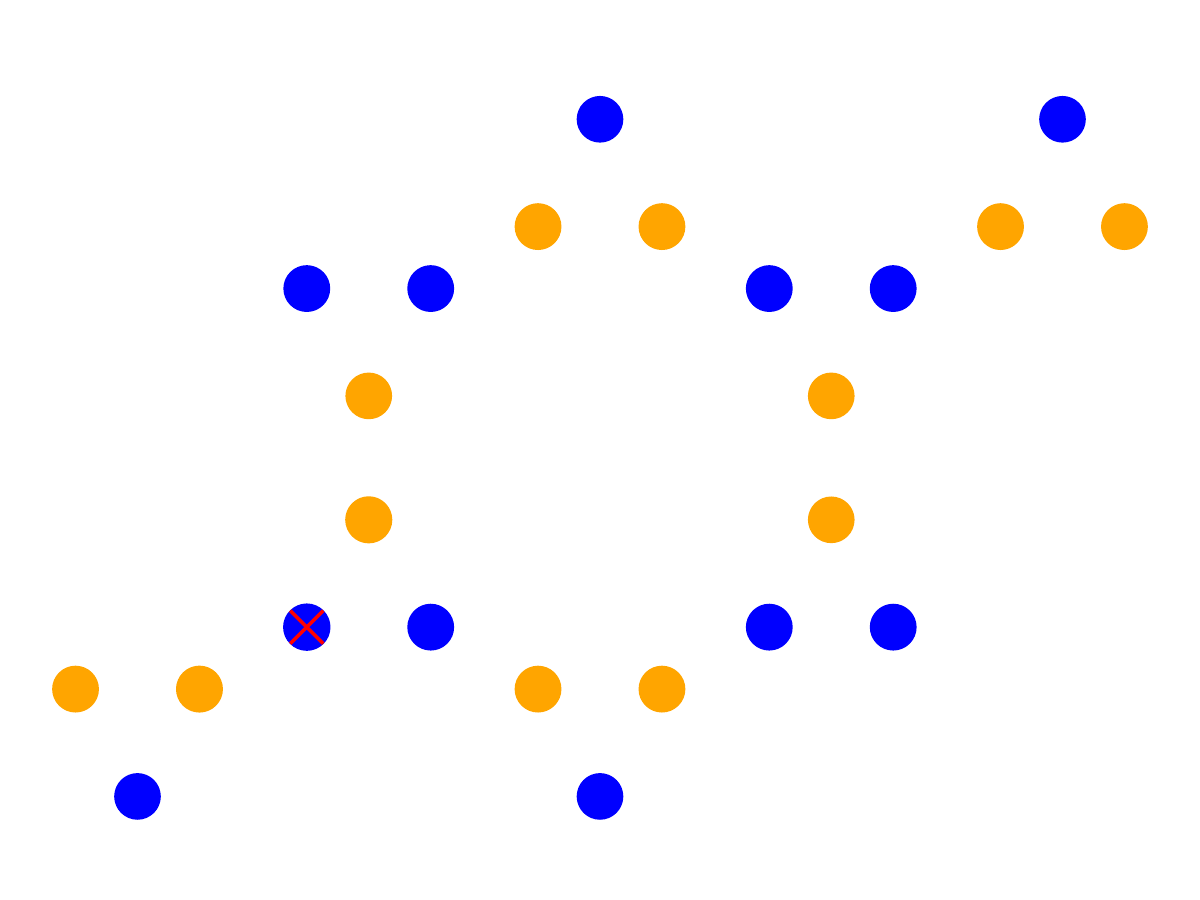}
   
  \caption{Left: The density-density correlations, $\left<n_i n_j\right>-\left<n_i\right>\left<n_j\right>$, obtained from exact diagonalization and plotted as a function of the position of site $j$ for the NI at $V_1=0.8$ and $V_2=2.4$, where the size of the dots represents the magnitude of the correlation and blue (orange) indicates positive (negative) sign. The reference site at $i$ is indicated by a red cross. The NI breaks $C_6$ rotation symmetry down to $C_3$, while preserving translational symmetry.  Right: The same but for the stripe phase at $V_1=20$ and $V_2=14$.  The stripe phase breaks translational symmetry by doubling the unit cell along the direction of unit vectors. 
  \label{fig_ninj}}
\end{figure}

  \section{Density-density correlations in the NI and stripe phases}\label{App:corr}
  Density-density correlations can also be used to reveal ordered phases.   Fig.~\ref{fig_ninj} plots the density-density correlation function for the NI and stripe phases obtained from ED.  Here we see agreement with the MFT calculation plotted in Fig.~2 of the main text.

 \section{Bond ordered phase}\label{App:BO}
\begin{figure}[t] 
  \resizebox{.8\columnwidth}{!}{\includegraphics{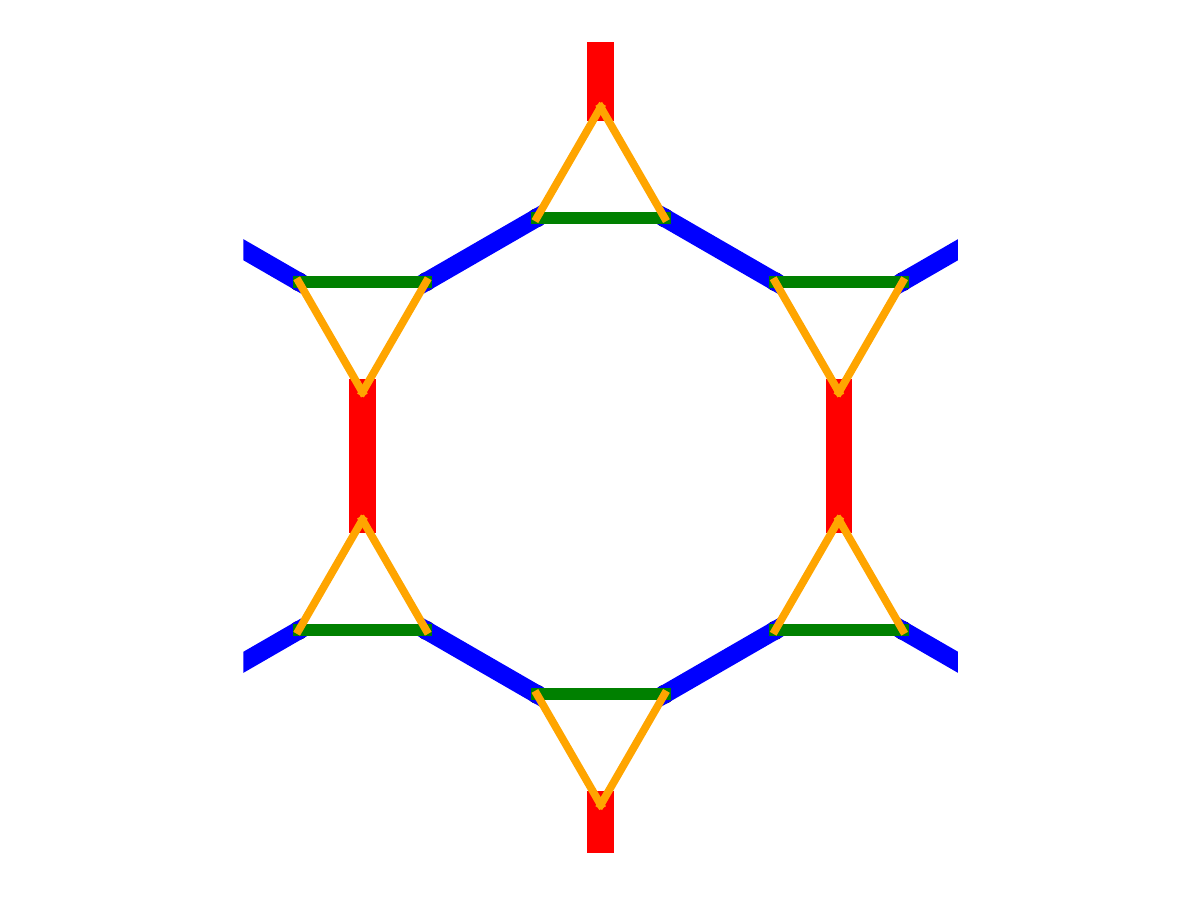}}  \caption{Bond average $\left<c_i^\dagger c_j\right>$ computed using exact diagonalization for an excitation of the BO phase at $V_1=5$ and $V_2=0.2$, where the bond averages are always real.  The width  of the line along a bond indicates the magnitude of the average and bonds with the same color have the same magnitude. This BO crystal excitation breaks rotational symmetry, but preserves translational symmetry. \label{fig_bond_order}}
\end{figure}

 The low $V_2$ part of the ED phase diagram, labeled BO in Fig.~3a of the main text, is distinct from MFT which instead finds a stripe phase. Within ED, the ground state here is non-degenerate and uniform with a massively degenerate excitation space.  We also find no currents in the ground state.  An example bond-average pattern for one of the excitations is shown in Fig.~\ref{fig_bond_order}.  Here we see that certain bonds have large tunneling but others are near zero.  A particle and hole form a bonding state along these bonds.

We interpret the uniform ground state as a unique superposition of excitations of the type depicted in  Fig.~\ref{fig_bond_order}.  The gap between the ground and excited states is near $t$.  We therefore conclude that tunneling superposes these excitations into a uniform BO state.

\bibliography{jabref_database_8_28_15.bib}
  
\end{document}